\documentclass[12pt,cite,epsf,epsfig]{article}
\usepackage{epsfig}

\begin{document}
\author{S. Dev\thanks{dev5703@yahoo.com},
Sanjeev Kumar\thanks{sanjeev3kumar@gmail.com}, Surender
Verma\thanks{ s\_7verma@yahoo.co.in} and Shivani
Gupta\thanks{shiroberts\_1980@yahoo.co.in}}
\title{CP Violation in Two Zero Texture Neutrino Mass Matrices}
\date{Department of Physics, Himachal Pradesh University,
Shimla 171005, INDIA}

\maketitle

\begin{abstract}
It has been shown that the neutrino mass matrices with two texture
zeros in the charged lepton basis predict non-zero 1-3 mixing and
are necessarily CP violating with one possible exception in class
C for maximal mixing.
\end{abstract}

There is no CP violation in the leptonic sector of the Standard
Model (SM) of fundamental particles and interactions. However, in
most extensions of SM, there can be several CP phases. In the
simplest three generation scenario, there is a Dirac type CP
violating phase in the leptonic mixing matrix. However, for
Majorana neutrinos, there could be two additional phases (Majorana
phases). It is possible to work in a parameterizations in which
all the three CP violating phases are situated in the charged
current leptonic mixing matrix. Without any loss of generality,
one can work in the flavor basis in which the charged lepton mass
matrix is diagonal so that the neutrino mass matrix carries all
the information about CP violation. In the flavor basis, the mass
matrix for Majorana neutrinos contains nine physical parameters
including the three mass eigenvalues, three mixing angles and the
three CP-violating phases. The two squared-mass differences
($\Delta m^2_{12}$ and $\Delta m^2_{13}$) and the two mixing
angles ($\theta_{12}$ and $\theta_{23}$) have been measured in
solar, atmospheric and reactor experiments. The third mixing angle
$\theta_{13}$ and the Dirac-type CP-violating phase $\delta$ are
expected to be measured in the forthcoming neutrino oscillation
experiments. The possible measurement of the effective Majorana
mass in neutrinoless double $\beta$ decay searches will provide an
additional constraint on the remaining three neutrino parameters
viz. the neutrino mass scale and two Majorana-type CP-violating
phases. While the neutrino mass scale will be independently
determined by the direct beta decay searches and cosmological
observations, the two Majorana phases will not be uniquely
determined from the measurement of effective Majorana mass even if
the absolute neutrino mass scale is known. Under the
circumstances, it is natural to employ other theoretical inputs
for the reconstruction of the neutrino mass matrix. The possible
forms of these additional theoretical inputs are limited by the
existing neutrino data. Several proposals have been made in the
literature to restrict the form of the neutrino mass matrix and to
reduce the number of free parameters which include presence of
texture zeros \cite{FGM paper, FGM papers, Tanimoto, a1a2 paper,
2t paper}, requirement of zero determinant \cite{Branco}, the zero
trace condition \cite{Zee} to name just a few. However, the
current neutrino oscillation data are consistent only with a
limited number of texture schemes \cite{FGM paper, FGM papers}. In
particular, the current neutrino oscillation data disallow all
neutrino mass matrices with three or more texture zeros in the
flavor basis. Out of the fifteen possible neutrino mass matrices
with two texture zeros, only seven are compatible with the current
neutrino oscillation data. The seven allowed two texture zero mass
matrices have been classified into three categories. The two class
A matrices of the types $A_1$ and $A_2$ give normal hierarchy of
neutrino masses. The class B matrices of types $B_1$, $B_2$, $B_3$
and $B_4$ yield a quasi-degenerate spectrum of neutrino masses.
The single class C matrix corresponds to inverted hierarchy of
neutrino masses.

The texture zeros in different positions of the neutrino mass
matrix can result from underlying flavor symmetries \cite{flavor
symmetries}. They can, also, be realized within the framework of
the see-saw mechanism \cite{see-saw textures}. Such textures can,
also, be obtained in the context of GUTs based on SO(10)
\cite{Goh;2003, Bando}. Recently, these texture zeros have been
derived from a simple flavor group $A_4$ or its $Z_3$ subgroup
\cite{A4}.

The existence of two texture zeros implies four real conditions to
be satisfied by the neutrino oscillation parameters. It is shown
that these four conditions cannot be satisfied if $\theta_{13}=0$.
The presence of texture zeros in the neutrino mass matrix,
further, demands certain cancellations induced by strong
correlations between CP-violating phases. It is shown that
neutrino mass matrices with two texture zeros in the charged
lepton basis must be necessarily CP-violating with one possible
exception in class C for $\theta_{23}=\frac{\pi}{4}$ and
$\theta_{13}=0$ where there will be no Dirac-type CP-violation and
we can, only, have Majorana-type CP-violation. This case can
become CP conserving for a special choice of $m_1$.

The neutrino mass matrix, $M$, can be parameterized in terms of
the three neutrino mass eigenvalues ($m_{1}$, $m_{2}$, $m_{3}$),
three neutrino mixing angles ($\theta_{12}$, $\theta_{23}$,
$\theta_{13}$) and one Dirac-type CP violating phase, $\delta$. If
neutrinos are Majorana particles, then, there are two additional
CP violating phases $\alpha$, $\beta$ in the neutrino mixing
matrix. The complex symmetric mass matrix $M$ can be diagonalized
by a complex unitary matrix $V$:
\begin{equation}
M=VM_{\nu}^{diag}V^{T}
\end{equation}
where $M_{\nu}^{diag}=Diag \{m_1,m_2,m_3\}$. The neutrino mixing
matrix $V$ can be written as
\begin{equation}
V\equiv U P=\left(
\begin{array}{ccc}
c_{12}c_{13} & s_{12}c_{13} & s_{13}e^{-i\delta} \\
-s_{12}c_{23}-c_{12}s_{23}s_{13}e^{i\delta} &
c_{12}c_{23}-s_{12}s_{23}s_{13}e^{i\delta} & s_{23}c_{13} \\
s_{12}s_{23}-c_{12}c_{23}s_{13}e^{i\delta} &
-c_{12}s_{23}-s_{12}c_{23}s_{13}e^{i\delta} & c_{23}c_{13}
\end{array}
\right)\left(
\begin{array}{ccc}
1 & 0 & 0 \\ 0 & e^{i\alpha} & 0 \\ 0 & 0 & e^{i(\beta+\delta)}
\end{array}
\right),
\end{equation}
where $s_{ij}=\sin\theta_{ij}$ and $c_{ij}=\cos\theta_{ij}$. The
matrix $V$ is called the neutrino mixing matrix or
Pontecorvo-Maki-Nakagawa-Sakata matrix. The matrix $U$ is the
lepton analogue of the Cabibbo-Kobayashi-Maskawa quark mixing
matrix and $P$ contains the two Majorana phases.

The elements of the neutrino mass matrix can be calculated from
Eq. (1). Some of the elements of $M$, which are equated to zero in
the various allowed texture zero schemes, are given by
\begin{equation}
M_{ee}=c_{13}^{2}c_{12}^{2}m_{1}+c_{13}^{2}s_{12}^{2}m_{2}e^{2i\alpha
}+s_{13}^{2}m_{3}e^{2i\beta },
\end{equation}
\begin{equation}
M_{e \mu}=c_{13}\{ s_{13}s_{23} e^{i \delta} ( e^{2 i \beta}
m_3-s^2_{12} e^{2 i \alpha} m_2 ) -c_{12}c_{23}s_{12} (m_1-e^{2 i
\alpha} m_2 ) -c^2_{12} s_{13} s_{23} e^{i \delta} m_1\},
\end{equation}
\begin{equation}
M_{e \tau}=c_{13}\{ s_{13}c_{23} e^{i \delta} ( e^{2 i \beta}
m_3-s^2_{12} e^{2 i \alpha} m_2 ) +c_{12}s_{23}s_{12} (m_1-e^{2 i
\alpha} m_2 ) -c^2_{12} s_{13} c_{23} e^{i \delta} m_1\},
\end{equation}
\begin{equation}
M_{\mu\mu}=m_1(c_{23}s_{12}+e^{i\delta}c_{12}s_{13}s_{23})^2+
e^{2i\alpha}m_2(c_{12}c_{23}-e^{i\delta}s_{12}s_{13}s_{23})^2+
e^{2i(\beta+\delta)}m_3c^2_{13}s^2_{23}
\end{equation}
and
\begin{equation}
M_{\tau\tau}=m_1(s_{23}s_{12}-e^{i\delta}c_{12}s_{13}c_{23})^2+
e^{2i\alpha}m_2(c_{12}s_{23}+e^{i\delta}s_{12}s_{13}c_{23})^2+
e^{2i(\beta+\delta)}m_3c^2_{13}c^2_{23}.
\end{equation}
The seven allowed neutrino mass matrices with two texture zeros
have been listed in Table 1.

\begin{table}[tb]
\begin{center}
\begin{tabular}{cc}
\hline
 Type  &\hspace{5cm}        Constraining Equations         \\
 \hline
 $A_1$ &\hspace{5cm}    $M_{ee}=0$, $M_{e\mu}=0$     \\
 $A_2$ &\hspace{5cm}    $M_{ee}=0$, $M_{e\tau}=0$     \\
 $B_1$ &\hspace{5cm}  $M_{e\tau}=0$, $M_{\mu\mu}=0$   \\
 $B_2$ &\hspace{5cm}  $M_{e\mu}=0$, $M_{\tau\tau}=0$  \\
 $B_3$ &\hspace{5cm}   $M_{e\mu}=0$, $M_{\mu\mu}=0$   \\
 $B_4$ &\hspace{5cm} $M_{e\tau}=0$, $M_{\tau\tau}=0$  \\
 $C$   &\hspace{5cm} $M_{\mu\mu}=0$, $M_{\tau\tau}=0$  \\
 \hline
\end{tabular}
\end{center}
\caption{Allowed two texture zero mass matrices.}
\end{table}

For neutrino mass matrices of type $A_1$, the conditions
$M_{ee}=0$ and $M_{e\mu}=0$ imply \cite{a1a2 paper}
\begin{equation}
s^2_{13} (m_1 c^2_{12} +m_2 s^2_{12} \cos 2\alpha-m_3 \cos 2
\beta)=c^2_{12} m_1+ s^2_{12} m_2 \cos 2 \alpha,
\end{equation}
\begin{equation}
s^2_{13} (m_2 s^2_{12} \sin 2\alpha-m_3 \sin 2 \beta)= s^2_{12}
m_2 \sin 2 \alpha,
\end{equation}
\begin{equation}
s_{13}(m_1 c^2_{12}\cos \delta +m_2 s_{12}^2 \cos
(2\alpha+\delta)-m_3\cos(2\beta+\delta))=\frac{c_{12}c_{23}s_{12}(m_2
\cos 2\alpha-m_1)}{s_{23}},
\end{equation}
\begin{equation}
s_{13}(m_1 c^2_{12}\sin \delta +m_2 s_{12}^2 \sin
(2\alpha+\delta)-m_3\sin(2\beta+\delta))=\frac{m_2
c_{12}c_{23}s_{12} \sin 2\alpha}{s_{23}}.
\end{equation}
It can be seen from Eqs. (9) and (11) that $\sin 2 \alpha$ must
vanish if $s_{13}$ is zero. However, the simultaneous solution of
Eqs. (8) and (10) under the condition $s_{13}=0$ implies that
$m_1=m_2=0$ which is inconsistent with solar neutrino data.
Therefore, $s_{13}$ should be non-zero for neutrino mass matrices
of type $A_1$ \cite{Tanimoto, a1a2 paper, 2t paper} which in turn
implies a non-vanishing $\sin 2 \alpha$ [Eqs. (9) and (11)]. It
can, also, be seen from Eqs. (9) and (11) that $\sin 2 \alpha=\sin
2 \beta=0$ if $\sin\delta=0$ which is not possible for non-zero
$s_{13}$. Similarly, one can show that $s_{13}$ and $\sin\delta$
are non-zero for neutrino mass matrices of type $A_2$. Hence,
neutrino mass matrices of class A are necessarily CP-violating
\cite{a1a2 paper}.

In category B, for neutrino mass matrices of type $B_3$, for
example, we have the conditions $M_{e\mu}=0$ and $M_{\mu\mu}=0$
which result in the following four real conditions:
\begin{equation}
s_{13}(m_1 c^2_{12}\cos \delta +m_2 s_{12}^2 \cos
(2\alpha+\delta)-m_3\cos(2\beta+\delta))=\frac{c_{12}c_{23}s_{12}(m_2
\cos 2\alpha-m_1)}{s_{23}},
\end{equation}
\begin{equation}
s_{13}(m_1 c^2_{12}\sin \delta +m_2 s_{12}^2 \sin
(2\alpha+\delta)-m_3\sin(2\beta+\delta))=\frac{m_2
c_{12}c_{23}s_{12} \sin 2\alpha}{s_{23}},
\end{equation}
\begin{equation}
s_{13}(m_2\cos(2\alpha+\delta)-m_1\cos\delta)=\frac{c^2_{23}(m_1
s^2_{12}+m_2c_{12}^2\cos 2\alpha
)+m_3s_{23}^2\cos2(\beta+\delta)}{c_{12}c_{23}s_{12}s_{23}}
\end{equation}
and
\begin{equation}
s_{13}(m_2\sin(2\alpha+\delta)-m_1\sin\delta)=
\frac{m_2c^2_{23}c_{12}^2\sin 2\alpha+m_3s_{23}^2\sin
2(\beta+\delta)}{c_{12}c_{23}s_{12}s_{23}}.
\end{equation}
It can be easily seen from Eqs. (12) and (13) that $m_1=m_2$ and
$\sin 2\alpha=0$ for $\theta_{13}=0$. Hence, $\theta_{13}$ and
$\alpha$ cannot be zero since solar mass squared difference
$\Delta m^2_{12}$ is non-zero. It follows from Eqs. (13) and (15)
that $\sin2\alpha$ and $\sin2\beta$ are zero if $\sin\delta=0$
which is not possible since $s_{13}$ is non-zero. Therefore,
$\sin\delta$ should be non-zero for neutrino mass matrices of type
$B_3$. Similarly, one can show that $\theta_{13}$ and $\sin\delta$
are non-zero for other neutrino mass matrices of class B viz.
$B_1$, $B_2$ and $B_4$. Thus, neutrino mass matrices of class B
are, also, necessarily CP violating.

The four real conditions to be satisfied by neutrino mass matrices
of class C can be written as
\begin{eqnarray}
s^2_{13} (m_3\cos 2 (\beta+\delta)-m_1 c^2_{12}\cos 2 \delta
-m_2s_{12}^2 \cos 2 (\alpha+\delta))  \nonumber\\
=m_1 s^2_{12}
+m_2c_{12}^2 \cos 2 \alpha+m_3\cos 2(\beta+\delta),
\end{eqnarray}
\begin{eqnarray}
s^2_{13} (m_3\sin 2 (\beta+\delta)-m_1 c^2_{12}\sin 2\delta
-m_2s_{12}^2 \sin 2 (\alpha+\delta)) \nonumber\\
=m_2c_{12}^2 \sin
2 \alpha+m_3\sin 2(\beta+\delta),
\end{eqnarray}
\begin{equation}
\sin 2\theta_{12} (m_1\cos\delta-m_2 \cos (2\alpha+\delta))
s_{13}=-2(m_1 s^2_{12}+m_2 c^2_{12} \cos 2\alpha)  \cot
2\theta_{23}
\end{equation}
and
\begin{equation}
\tan\theta_{12}(m_1\sin\delta-m_2 \sin (2\alpha+\delta))
s_{13}=-m_2 \sin 2\alpha\cot 2\theta_{23}.
\end{equation}
When, $\theta_{23}$ is non-maximal, the condition $s_{13}=0$ gives
$\sin2\alpha=0$ and $\frac{m_2}{m_1}=\tan^2\theta_{12}<1$ from
Eqs. (18) and (19) which contradicts the solar mass hierarchy. It
can, also, be shown that the above system of equations cannot
simultaneously hold for $\sin\delta=0$ since Eq. (19) implies that
$\sin 2\alpha=0$ under this condition which, again, gives
$s_{13}=0$ which, as discussed above, is inconsistent with the
data. Therefore, the neutrino mass matrices of class C are
necessarily CP violating for non-maximal 2-3 mixing.

For maximal mixing, $\theta_{23}=\frac{\pi}{4}$ and the Eqs. (18)
and (19) imply that $\theta_{13}=0$. In this case, there will be
no Dirac-type CP violation since $s_{13}=0$ and we are left with
the Eqs. (16) and (17) only which can be written as
\begin{equation}
s^2_{12} m_1+c^2_{12}m_2 \cos 2\alpha=-m_3 \cos 2\beta,
\end{equation}
and
\begin{equation}
c^2_{12}m_2 \sin 2\alpha=-m_3 \sin 2\beta.
\end{equation}
From Eqs. (20) and (21), the only possible CP conserving solution
for neutrino mass matrices of class C with maximal mixing is given
by
\begin{equation}
\alpha=\frac{\pi}{2},~\beta=0
\end{equation}
and
\begin{equation}
m_1^2=\frac{1}{\sin^2 2 \theta_{12}}\frac{(\Delta
m^2_{12}c^4_{12}-\Delta m^2_{13})^2}{\Delta
m^2_{12}c^2_{12}-\Delta m^2_{13}}.
\end{equation}
This solution is valid only for inverted hierarchy with $\Delta
m^2_{13}$ negative.  However, it may be difficult to realize the
relationship given in Eq. (23) in a realistic model of lepton
masses and mixings. In this special CP conserving case, the CP
parity of $\nu_2$ is odd and that of $\nu_3$ is even. CP is
conserved in the region allowed by Eqs. (22) and (23) within the
complete parameter space allowed by Eqs. (20) and (21) and in the
rest of the allowed parameter space, the neutrino mass matrices of
class C with maximal mixing will have CP violation of Majorana
type only. This is a consequence of the fact that the third
neutrino mass eigenstate $\nu_3$ gets decoupled from the first two
mass eigenstates $\nu_1$ and $\nu_2$ for vanishing $s_{13}$ and
the remaining $2 \times 2$ sub-matrix can have only Majorana-type
CP violation with a single physical Majorana phase. Hence, the
neutrino mass matrices of class C with maximal mixing can be CP
conserving only for $m_1$ given by Eq. (23). For all other values
of $m_1$, the neutrino mass matrices of class C with maximal
mixing will exhibit CP violation with one physical Majorana phase.

It should, also, be noted that the neutrino mass matrices of class
C with maximal mixing are $\mu-\tau$ symmetric since we have
\begin{equation}
M_{e\mu}=-M_{e\tau} ~ and ~ M_{\mu\mu}=M_{\tau\tau}=0
\end{equation}
in this special case. However, the neutrino mass matrices of
classes A and B can never become $\mu-\tau$ symmetric which
results in a different CP structure of class C as compared to that
of classes A and B of neutrino mass matrices.

In conclusion, it has been shown that CP violation is inherent in
the two texture zero scheme of Frampton, Glashow and Marfatia
\cite{FGM paper} and existence of texture zeros in the neutrino
mass matrix may provide useful hints for unraveling the dynamics
of the CP violation. The existence of the texture zeros in the
neutrino mass matrix in the flavor basis could result from a
certain flavor symmetry. This, in turn, would require non-zero
values of the mixing angle $\theta_{13}$ and the CP violating
phases to ensure the desired cancellations in the specific
elements of the neutrino mass matrix.

\vspace{1cm}
\begin{large}
\textbf{Acknowledgments} \end{large}

The research work of S. D. and S. V. is supported by the Board of
Research in Nuclear Sciences (BRNS), Department of Atomic Energy,
Government of India \textit{vide} Grant No. 2004/ 37/ 23/ BRNS/
399. S. K. acknowledges the financial support provided by Council
for Scientific and Industrial Research (CSIR), Government of
India.

\end{document}